\documentclass[12pt]{article}
\usepackage{graphicx}
\usepackage{graphicx}				

\newcommand\pubnumber{}
\newcommand\pubdate{\today}

\def\support{\footnote{Now at the Centro Cient\'{i}fico Tecnol\'{o}gico de Valpara\'{i}so \\
Universidad T\'{e}cnica Federico Santa Mar\'{i}a \\ 
Valpara\'{i}so, Chile}}
          
\usepackage{amssymb}
\usepackage{fleqn}
\usepackage{graphicx}
\usepackage{dcolumn}
\usepackage{amsmath}
\usepackage{epsfig}

\RequirePackage{xspace}

\input{babarsym}

\def\babar{\mbox{\slshape B\kern-0.1em{\smaller A}\kern-0.1em
    B\kern-0.1em{\smaller A\kern-0.2em R}}}
\def\ie{{\it i.e.}}

\def\Dpm    {\ensuremath{D^{\pm}}\xspace}
\def\Dp    {\ensuremath{D^{+}}\xspace}
\def\Dm    {\ensuremath{D^{-}}\xspace}
\def\Dkkpi    {\ensuremath{D^{+}\to K^{+}K^{-}\pi^{+}}\xspace}

\def\kk       {\ensuremath{K^{+}K^{-}}\xspace}
\def\kpi      {\ensuremath{K^{-}\pi^{+}}\xspace}
\def\mkk      {\ensuremath{m(K^{+}K^{-})}\xspace}
\def\mkpi     {\ensuremath{m(K^{-}\pi^{+})}\xspace}

\def\coscm    {\ensuremath{\cos(\theta_{\mathrm{CM}})}\xspace}

\def\msqkk    {\ensuremath{m^{2}(K^{+}K^{-})}\xspace}
\def\msqkpi   {\ensuremath{m^{2}(K^{-}\pi^{+})}\xspace}
\def\CP       {\ensuremath{C\!P}\xspace}

\def\CPV      {\ensuremath{C\!PV}\xspace}

\def\etal     {{\it et al.}}
\def\Kres     {\ensuremath{\bar{K}^*(892)^0}}
\def\phires   {\ensuremath{\phi(1020)}}

\newcommand{\gevcccc}{\ensuremath{{\mathrm{\,Ge\kern -0.1em V^2\!/}c^4}}\xspace}

\def\Title#1{\begin{center} {\Large #1 } \end{center}}
\def\Author#1{\begin{center}{ \sc #1} \end{center}}
\def\Address#1{\begin{center}{ \it #1} \end{center}}

\newcommand\pubblock{\rightline{\begin{tabular}{l} \pubnumber\\
         \pubdate  \end{tabular}}}
\newenvironment{Abstract}{\begin{quotation}  }{\end{quotation}}
\newenvironment{Presented}{\begin{quotation} \begin{center} 
             PRESENTED AT\end{center}\bigskip 
      \begin{center}\begin{large}}{\end{large}\end{center} \end{quotation}}
\def\Acknowledgements{\bigskip  \bigskip \begin{center} \begin{large}
             \bf ACKNOWLEDGEMENTS \end{large}\end{center}}

\def\beq{\begin{equation}}
\def\eeq#1{\label{#1}\end{equation}}
\def\eeqn{\end{equation}}

\def\beqa{\begin{eqnarray}}
\def\eeqa#1{\label{#1}\end{eqnarray}}
\def\eeqan{\end{eqnarray}}

\let\bar=\overbar

\def\etal{{\it et al.}}
\def\ie{{\it i.e.}}

\def\Dslash{\not{\hbox{\kern-4pt $D$}}}
\def\dslash{\not{\hbox{\kern-2pt $\del$}}}

\def\BR{\mbox{\rm BR}}

\def\msb{{\bar{\ssstyle M \kern -1pt S}}}

\begin{document}
\begin{titlepage}
\pubblock

\vfill
\Title{Search for \CP Violation in Charm at $e^+e^-$ colliders}
\vfill
\Author{Ryan Mackenzie White\support}
\Address{Department of Physics \\
University of South Carolina, USA
}
\vfill
\begin{Abstract}
In this proceeding, I discuss results from the \babar~and Belle collaborations for
searches of direct \CP violation in the singly Cabibbo-suppressed decay $\Dpm\to\kk\pi^\pm$ 
from \epem annihilation data collected at a center-of-mass energy at or just below the \Y4S
resonance. The Belle collaboration measures the \CP asymmetry as a function of
the production angle of the \Dpm meson in the quasi two-body  $\Dpm\to\phi\pi^\pm$
decay. The \babar~experiment studies the entire phase-space with 
model-independent and
model-dependent Dalitz plot analysis techniques to search for
\CP-violating asymmetries in the various intermediate states, in addition to a phase-space
 integrated measurement
as a function of the production angle. No evidence for \CP violation is
reported from either experiment.
\end{Abstract}
\vfill
\begin{Presented}
2013 Flavor Physics and CP Violation (FPCP-2013) \\
Buzios, Rio de Janeiro, Brazil, May 19-24 2013, 11 pages, 4 figures
\end{Presented}
\vfill
\end{titlepage}
\def\thefootnote{\fnsymbol{footnote}}
\setcounter{footnote}{0}

\section{Introduction}
Searches for \CP violation ({\CPV}) in charm meson decays provide a
probe of physics beyond the Standard Model (SM).  Singly
Cabibbo-suppressed (SCS) decays can exhibit direct \CP asymmetries due
to interference between tree-level transitions and $|\Delta C| = 1$
penguin-level transitions if there is both a strong and a weak phase
difference between the two amplitudes. In the SM, the resulting
asymmetries are suppressed by
$\mathcal{O}(|V_{cb}V_{ub}/V_{cs}V_{us}|)\sim10^{-3}$, where
$V_{\mathrm{ij}}$ are elements of the Cabibbo-Kobayashi-Maskawa
quark-mixing matrix~\cite{CKM}. A larger measured value of the \CP
asymmetry could be a consequence of the enhancement of penguin
amplitudes in $D$ meson decays due to final-state
interactions~\cite{GronauRosner,ChengChiang}, or of new
physics~\cite{GKN07,AMP08}.

SCS three-body decays of charm mesons are dominated by quasi-two-body 
decays with resonant intermediate states. Direct \CP violation (\CPV) can be localized in a 
particular region of phase-space, and the final-state interactions in these decays may enhance small new physics 
\CP phases. Analysis techniques which utilize the information contained in the Dalitz plot
take advantage of the dynamics in these decays to probe for \CPV.
Both the \babar~and Belle collaborations have analyzed the decay \Dkkpi~\cite{cc}
using several complimentary analysis techniques to measure \CP violating asymmetries
as a function of the Dalitz plot position with sufficient precision to probe 
for new physics~\cite{BabarKKpi,BellePhiPi}.

The Belle collaboration studied \CP asymmetries in charged $D^+\rightarrow\phi\pi^+$ and 
$D_s^+\rightarrow\phi\pi^+$ decays using 955 \invfb of data recorded with the Belle detector at the KEKB asymmetric energy \epem collider. The study of \Dkkpi from \babar~collaboration 
include a measurement of the
integrated \CP asymmetry, a comparison of the binned $D^+$ and $D^-$ Dalitz plots, a
comparison of the Legendre polynomial moment distributions for the \kk
and \kpi systems, and a comparison of parameterized fits to the Dalitz
plots. The analysis is based on a sample of 476 \invfb of data collected with the
\babar~detector at the SLAC \pep2
collider.

The production of \Dp (and \Dm) mesons from the $e^+e^-\to c\bar{c}$ process
is not symmetric in \coscm; this forward-backward (FB) asymmetry,
coupled with the asymmetric acceptance of the detector, results in
different yields for \Dp and \Dm events. The FB asymmetry, to first
order, arises from the interference of the separate annihilation
processes involving a virtual photon and a $Z^{0}$ boson. The
charge asymmetry $A$ in a given interval of \coscm by
\begin{equation}
  A(\coscm) \equiv \frac{ N_{D^{+}}/\epsilon_{D^{+}}  - N_{D^{-}}/\epsilon_{D^{-}}}
    {N_{D^{+}}/\epsilon_{D^{+}} + N_{D^{-}}/\epsilon_{D^{-}}},
\end{equation}
where $N_{D^\pm}$ and $\epsilon_{D^\pm}$ are the yield and efficiency,
respectively, in the given \coscm bin. 
The FB asymmetry is removed by averaging $A$ over intervals
symmetric in \coscm, \ie, by evaluating
\begin{equation}
  \label{eqn:acp}
  A_{CP} \equiv \frac{A(\coscm) + A(-\coscm)}{2}.
\end{equation}
In contrast to the \babar~experiment, Belle utilizes the Cabibbo-favored (CF) $D_s^+\rightarrow\phi\pi^+$ decay and 
measures
\begin{equation}
 \Delta A_{rec} = \frac{ N_{D^{+}}  - N_{D^{-}}}
    {N_{D^{+}} + N_{D^{-}}} -  \frac{ N_{D_s^{+}}  - N_{D_s^{-}}}
    {N_{D_s^{+}} + N_{D_s^{-}}}
    \end{equation}
    
in order to cancel detector-induced asymmetries and other systematic effects. This CF 
decay is governed by the CKM matrix elements $V_{cs}V^*_{ud}$ and is expected to have negligible
CP asymmetry~\cite{BN99}, therefore, a measurement of $ \Delta A_{rec}$ probes $A_{CP}^{D^+\rightarrow\phi\pi^+}$.

Both the \babar~and Belle experiment find no evidence for direct CP violation measured as a function
of \coscm. The Belle experiment measures
\begin{equation}
A_{CP}^{D^+\rightarrow\phi\pi^+} = (+0.51 \pm 0.28 \pm 0.05)\%
\end{equation}
and the \babar~measurement integrated over the entire phase-space is  
\begin{equation}
A_{CP}^{\Dkkpi} = (+0.37\pm0.30\pm0.15)\%.
\end{equation}

In  order to probe for \CPV as a function of position of the Dalitz plot, the \babar~experiment measures 
the asymmetry in intervals of the Dalitz plot either as a function of (\msqkk,\msqkpi), \mkk or \mkpi, 
as well as differences between the magnitudes and phase angles of resonance and non-resonant amplitudes
contributing to the the decay.

\begin{figure}[!tb]
\begin{center}
\includegraphics[width=0.45\textwidth]{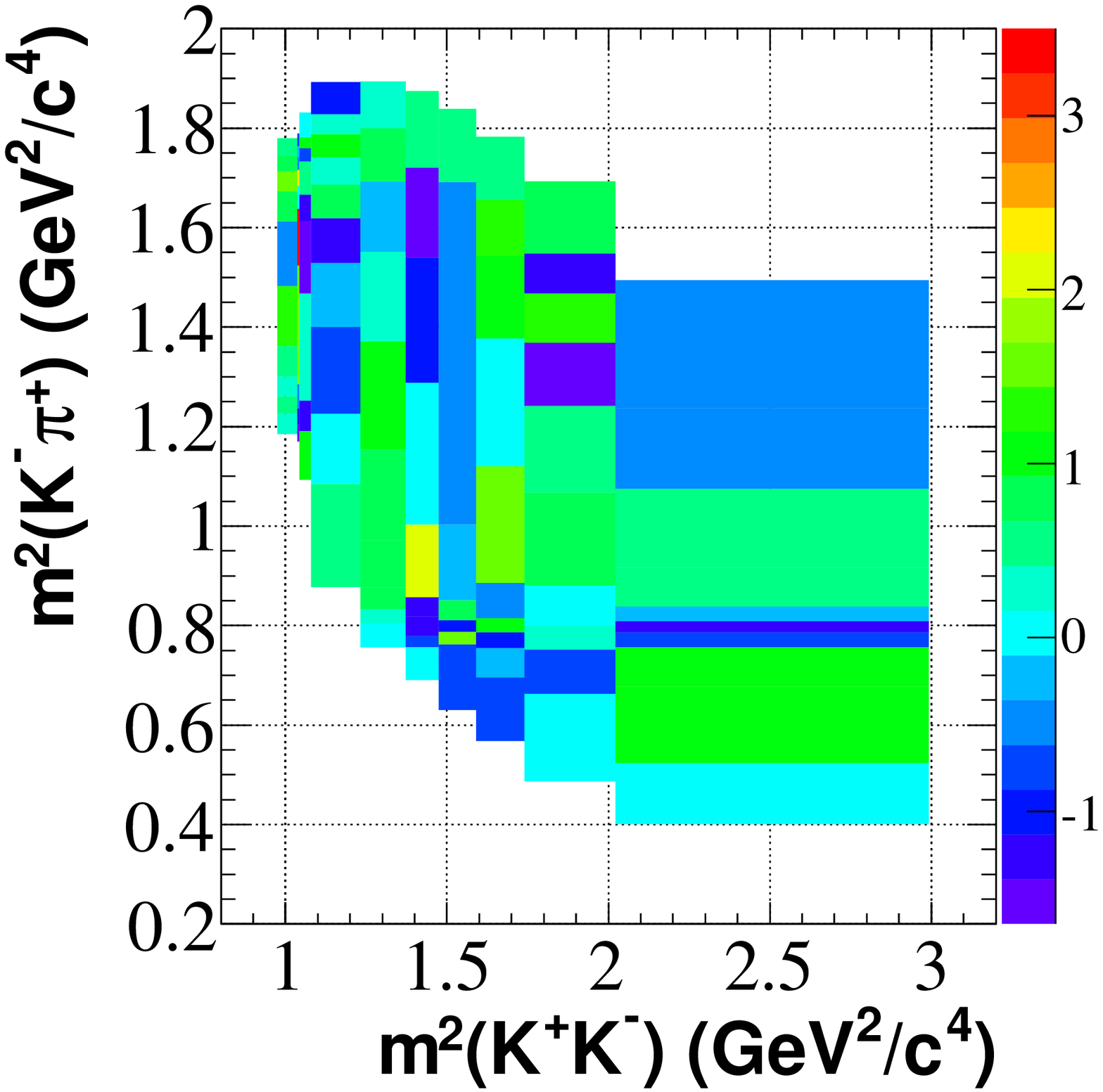}
\includegraphics[width=0.45\textwidth]{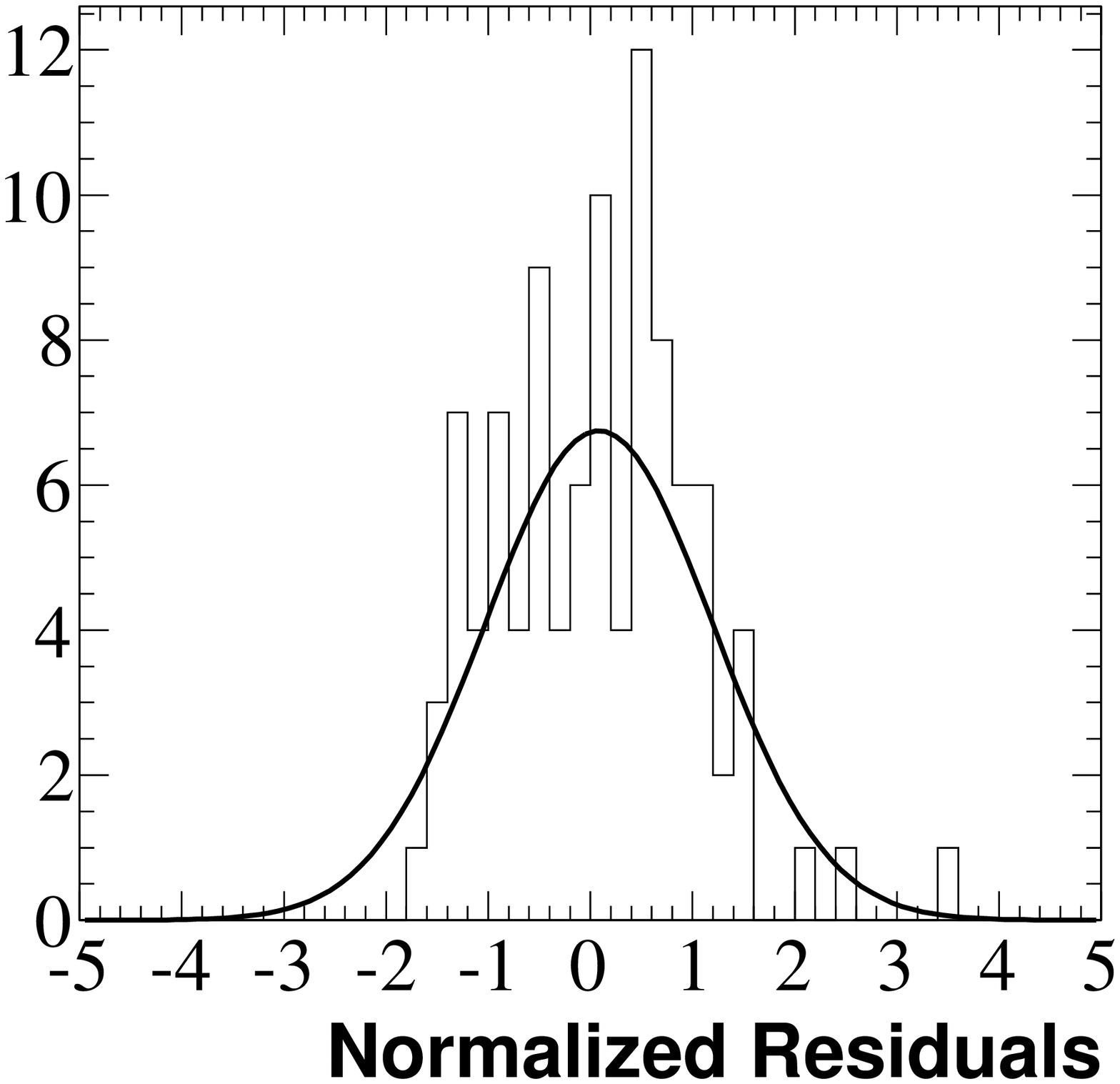}
\vspace{-0.3cm}
\caption{Normalized residuals of the $D^+$ and $D^-$
  Dalitz plots in equally populated intervals (top) and their
  distribution fitted with a Gaussian function (bottom).}
\label{fig:dpresasym}
\vspace{-0.7cm}
\end{center}
\end{figure}

To search for {\CPV} in intervals of \msqkk versus \msqkpi, \babar~measures
normalized residuals $\Delta$ for the efficiency-corrected and
background-subtracted \Dp and \Dm Dalitz
plots, where $\Delta$ is defined by
\begin{equation}
  \Delta \equiv {n(D^+) - Rn(D^-) \over \sqrt{\sigma^2(D^+) + R^2\sigma^2(D^-)}},
\end{equation}
with $n(\Dp)$ and $n(\Dm)$ the observed number of \Dp and \Dm mesons in
an interval of the Dalitz plot, where $\sigma(\Dp)$ and $\sigma(\Dm)$
are the corresponding statistical uncertainties.  $R$ is the ratio of efficiency-corrected
yields of \Dp and \Dm. The results for
$\Delta$ are shown in Fig.~\ref{fig:dpresasym}.  Note that the intervals
for Fig.~\ref{fig:dpresasym} are adjusted so that each interval contains
approximately the same number of events.  \babar~calculates the quantity
$\chi^2/(\nu-1) = (\sum_{i=1}^{\nu}\Delta^2)/(\nu - 1)$, where $\nu$ is
the number of intervals in the Dalitz plot. \babar~fits the distribution of
normalized residuals to a Gaussian function, whose mean and
root-mean-squared (RMS) deviation values are consistent with
zero and one, respectively.  The $\chi^2 = 90.2$ for 100 intervals
with a Gaussian residual mean of 0.08 $\pm$ 0.15, RMS deviation of 1.11
$\pm$ 0.15, and a consistency at the 72$\%$ level that the Dalitz plots
do not exhibit \CP asymmetry.

The Legendre polynomial moments of the cosine of the helicity angle of
the \Dpm decay products reflect the spin and mass of the intermediate
resonant and nonresonant amplitudes, and the interference effects among
them~\cite{Aubert:2007kkpi0}.  A comparison of these moments between the
\Dp and \Dm two-body mass distributions provides a model-independent
method to search for \CP violation in the Dalitz plot, and to study its
mass and spin structure. The helicity angle $\theta_H$ for
decays $\Dp\to(r\to K^+K^-)\pi^+$ via resonance $r$ is defined as the angle between
the $K^+$ direction in the $K^+K^-$ rest frame and the prior direction
of the $K^+K^-$ system in the \Dp rest frame. For decays $\Dp\to(r\to
K^-\pi^+)K^+$ via resonance $r$, $\theta_H$ is defined as the angle
between the $K^-$ direction in the $K^-\pi^+$ system and the prior
direction of the $K^-\pi^+$ system in the \Dp rest frame.

\begin{figure}[!tb]
\begin{center}
\includegraphics[width=0.45\textwidth]{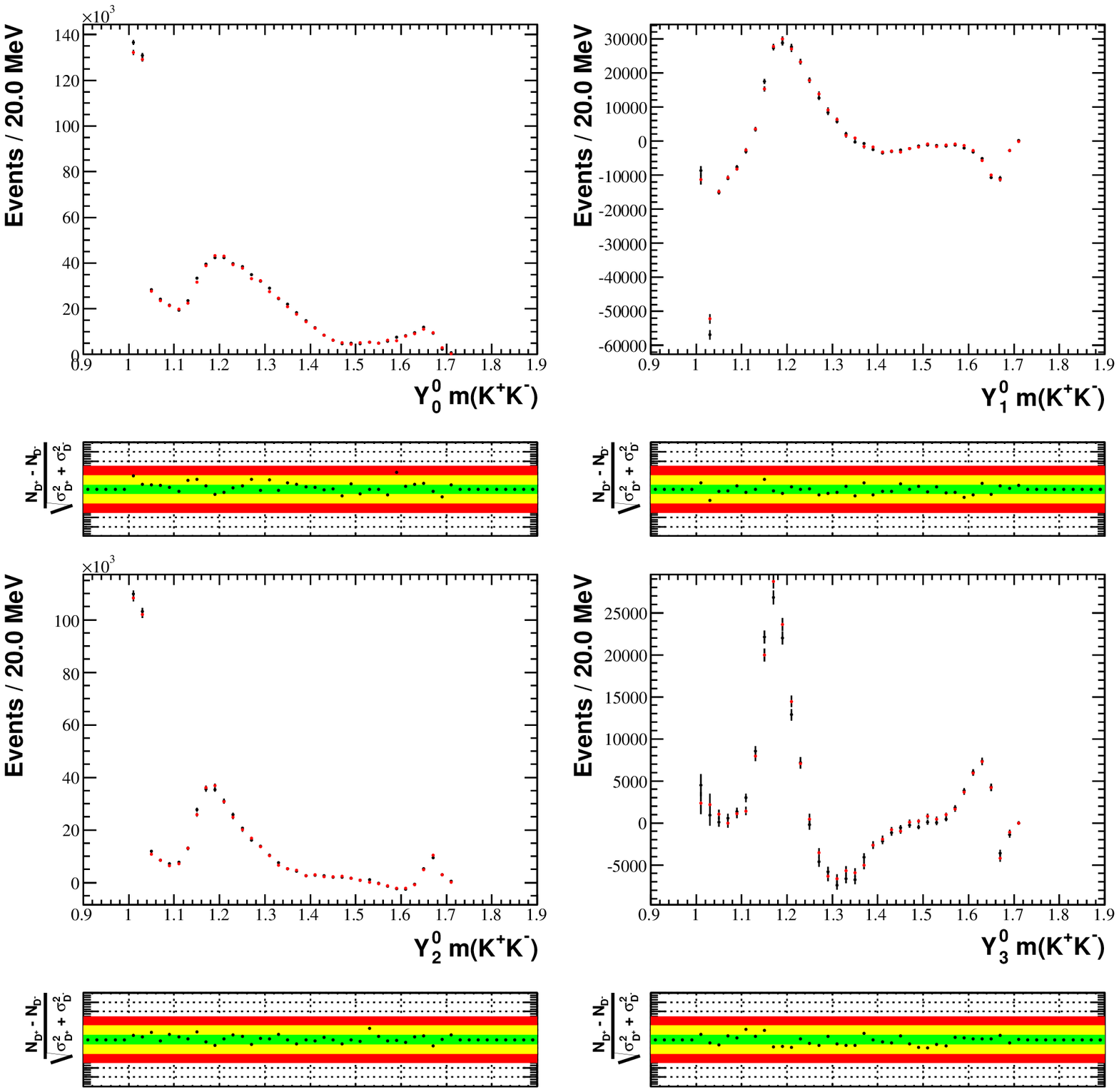}
\includegraphics[width=0.45\textwidth]{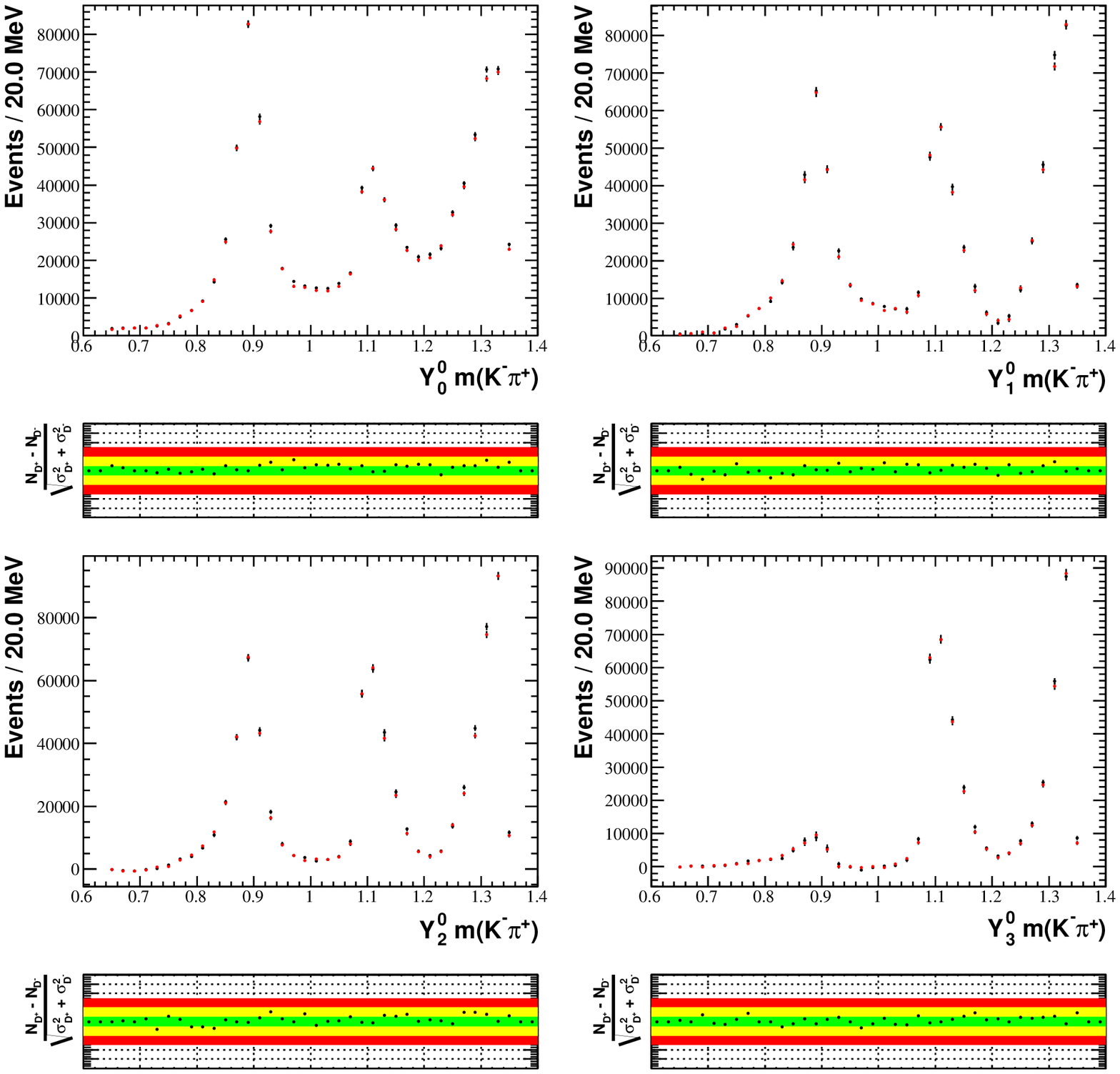}
\vspace{-0.3cm}
\caption{\Dp (black) and \Dm (red) Legendre polynomial moment distributions up to $l=3$ for 
the efficiency-corrected and background-subtracted two-body mass distributions
of \mkk (left) and \mkpi (right) and the normalized residuals $X_i^l$.}
\label{fig:moments}
\vspace{-0.7cm}
\end{center}
\end{figure}

The Legendre polynomial moment distribution for order $l$ is defined as
the efficiency-corrected and background-subtracted invariant two-body
mass distribution \mkk or \mkpi, weighted by the spherical harmonic
$Y_l^0[\cos(\theta_H)]=\sqrt{2l+1/4\pi}P_l[\cos(\theta_H)]$, where $P_l$
is the Legendre polynomial (Fig.~\ref{fig:moments}). The two-body invariant mass
interval weight is defined as $W_i^{(l)}\equiv{(\sum_j w_{ij}^{(l)S} - \sum_k
w_{ik}^{(l)B})} / {\langle\epsilon_i\rangle}$, where
$w_{ij}^{(l)}(w_{ik}^{(l)})$ is the value of $Y_l$ for the
$j^{\mathrm{th}}(k^{\mathrm{th}})$ event in the $i^{\mathrm{th}}$ interval
and $\langle\epsilon_i\rangle$ is the average efficiency for the
$i^{\mathrm{th}}$ interval. The superscripts $S$ and $B$ refer to the signal
and background components, respectively.  The uncertainty on $W_i^{(l)}$
is ${\sigma^{(l)}}\equiv\sqrt{{\sum_j (w_{ij}^{(l)S})^2 + \sum_k
(w_{ik}^{(l)B})^2}/{\langle\epsilon_i\rangle^2}}$.  To study differences
between the \Dp and \Dm amplitudes, the quantities $X_i^l$
for $l$ ranging from zero to seven in a two-body invariant mass
interval, where
\begin{equation}
  X_i^l = \frac{(W_i^{(l)}(\Dp) - RW_i^{(l)}(\Dm))}{\sqrt{{\sigma_i^{(l)}}^2(\Dp) + R^2{\sigma_i^{(l)}}^2(\Dm)}}.
\end{equation}
are calculated.

\babar~calculates the $\chi^2/$ndof over 36 mass intervals in the \kk and
\kpi moments using
\begin{equation}
  \chi^2 = \sum_i\sum_{l_1}\sum_{l_2} X_i^{(l_1)} \rho_i^{l_1l_2} X_i^{(l_2)},
\end{equation}
where $\rho_i^{l_1l_2}$ is the correlation coefficient between $X^{l_1}$ and $X^{l_2}$,
\begin{equation}
\rho_i^{l_1 l_2} \equiv 
\frac{ \langle X_i^{(l_1)} X_i^{(l_2)} \rangle - 
    \langle X_i^{(l_1)}\rangle\langle X_i^{(l_2)}\rangle}
{
    \sqrt{ \langle {X_i^{(l_1)}}^2 \rangle - {\langle X_i^{(l_1)} \rangle}^2} 
    \sqrt{ \langle {X_i^{(l_2)}}^2 \rangle - {\langle X_i^{(l_2)} \rangle}^2} 
},
\end{equation}
and where the number of degrees of freedom (ndof) is given by the
product of the number of mass intervals and the number of moments, minus
one due to the constraint that the overall rates of \Dp and \Dm mesons
be equal.  The $\chi^2/$ndof is found to be 1.10 and 1.09 for the \kk and
\kpi moments, respectively (for ndof $= 287$), which corresponds to a
probability of 11\% and 13\%, again respectively, for the null
hypothesis (no \CPV).

The Dalitz plot amplitude ${\mathcal A}$ can be described by an isobar
model, which is parameterized as a coherent sum of amplitudes for a set
of two-body intermediate states $r$. Each amplitude has a complex
coefficient, \ie, 
\begin{equation}
{\mathcal A}_r[\msqkk,\msqkpi]=\sum_r{\mathcal
M}_re^{i\phi_r}F_r[\msqkk,\msqkpi],
\end{equation}
where ${\mathcal M}_r$ and $\phi_r$ are real numbers, and the $F_r$ are
dynamical functions describing the intermediate resonances~\cite{CLEO-spin, Asner, BWtext}. The complex
coefficient may also be parameterized in Cartesian form, $x_r =
{\mathcal M}_r\cos\phi_r$ and $y_r = {\mathcal M}_r\sin\phi_r$.  The \Kres~ is chosen 
as the reference amplitude in the \CP-symmetric and
\CP-violating fits to the data, such that ${\mathcal M}_{\Kres} = 1$ and
$\phi_{\Kres} = 0$.

The \CP conserving background is modeled using events from the sideband regions 
of the \Dp mass distribution, which is comprised of the \Kres~and
\phires~resonance contributions and combinatorial background. The
combinatorial background outside the resonant regions has a smooth shape
and is modeled with the non-parametric $k$-nearest-neighbor density
estimator~\cite{knn}. The \Kres~and \phires~regions are composed of
the resonant structure and a linear combinatorial background, which is
parameterized as a function of the two-body mass and the cosine of the
helicity angle. The model consists of a Breit-Wigner (BW) PDF to
describe the resonant line shape, and a first-order polynomial in mass
to describe the combinatorial shape. These are further multiplied by a
sum over low-order Legendre polynomials to model the angular dependence.

\begin{table}[!ht]
\begin{center}
\caption{Resonance mass and width values determined from the isobar model fit to the combined Dalitz-plot distribution.}
\setlength{\extrarowheight}{2pt}
\begin{tabular}{lcc}
\hline
\hline
Resonance & Mass (MeV/c$^2$) & Width (MeV)  \\
\hline
$\bar{K}^{*}(892)^{0}$ & 895.53 $\pm$ 0.17 & 44.90 $\pm$ 0.30 \\
$\phi(1020)$ & 1019.48 $\pm$ 0.01 & 4.37 $\pm$  0.02 \\
$a_{0}(1450)$ & 1441.59 $\pm$ 3.77 & 268.58 $\pm$ 5.28 \\
$\bar{K}^{*}_{0}(1430)^{0}$ & 1431.88 $\pm$ 5.89 & 293.62 $\pm$ 3.83 \\
$\bar{K}^{*}(1680)^{0}$  & 1716.88 $\pm$ 21.03 & 319.28 $\pm$ 109.07  \\
$f_{0}(1370)$ & 1221.59 $\pm$ 2.46 & 281.48 $\pm$ 6.6\\
$\kappa(800)$ & 798.35 $\pm$ 1.79 & 405.25 $\pm$ 5.05\\
\hline
\hline
\end{tabular} 
\label{tab::fit_resparms}
\end{center}
\end{table}

Assuming no {\CPV}, an unbinned maximum-likelihood fit is performed to
determine the relative fractions for the resonances contributing to the
decay: $\bar{K}^*(892)^0$, $\bar{K}^*(1430)^0$, $\phi(1020)$,
$a_{0}(1450)$, $\phi(1680)$, $\bar{K}^*_2(1430)^0$, $\bar{K}^*(1680)^0$,
$\bar{K}^*_1(1410)^0$, $f_2(1270)$, $f_0(1370)$, $f_0(1500)$,
$f_2^{\prime}(1525)$, $\kappa(800)$, $f_0(980)$, $f_0(1710)$, and a
nonresonant (NR) constant amplitude over the entire Dalitz plot.
The negative log likelihood (NLL) function is minimized
\begin{eqnarray}
-2 \ln \mathcal{L} = \qquad \qquad \qquad \qquad \qquad \qquad \qquad \qquad 
        \nonumber \\ -2 \sum_{i=1}^{N} \ln \bigg[ p(m_{i})
	\frac{\epsilon_{\mathrm{MC}}(x_1,x_2) S(x_1,x_2)} 
	{\iint\epsilon_{\mathrm{MC}}(x_1,x_2) S(x_1,x_2)\mathrm{d}x_1\mathrm{d}x_2} 
	+ \nonumber\\ ( 1 - p(m_{i}) ) 
	\frac{B(x_1,x_2)}{\iint B(x_1,x_2)\mathrm{d}x_1\mathrm{d}x_2}  \bigg ],\qquad
\end{eqnarray}
where $N$ is the number of events.
The reconstructed $D^{+}$ mass-dependent probability $p(m)$ is defined
as $p(m_{i}) = \frac{S(m_{i})}{S(m_{i})+B(m_{i})}$, where $S(m)$ and
$B(m)$ are the signal and background PDFs, whose parameters are determined 
from a fit of the \Dp mass distribution described; $x_1=\msqkk$ and $x_2=\msqkpi$, $S(x_1,x_2)$ is the
Dalitz plot amplitude-squared, $\epsilon_{\mathrm{MC}}$ is the ANN efficiency, 
and $B(x_1,x_2)$ is the \CP-symmetric background PDF.

\begin{figure*}[!tb]
\begin{center}
\includegraphics[width=0.98\textwidth]{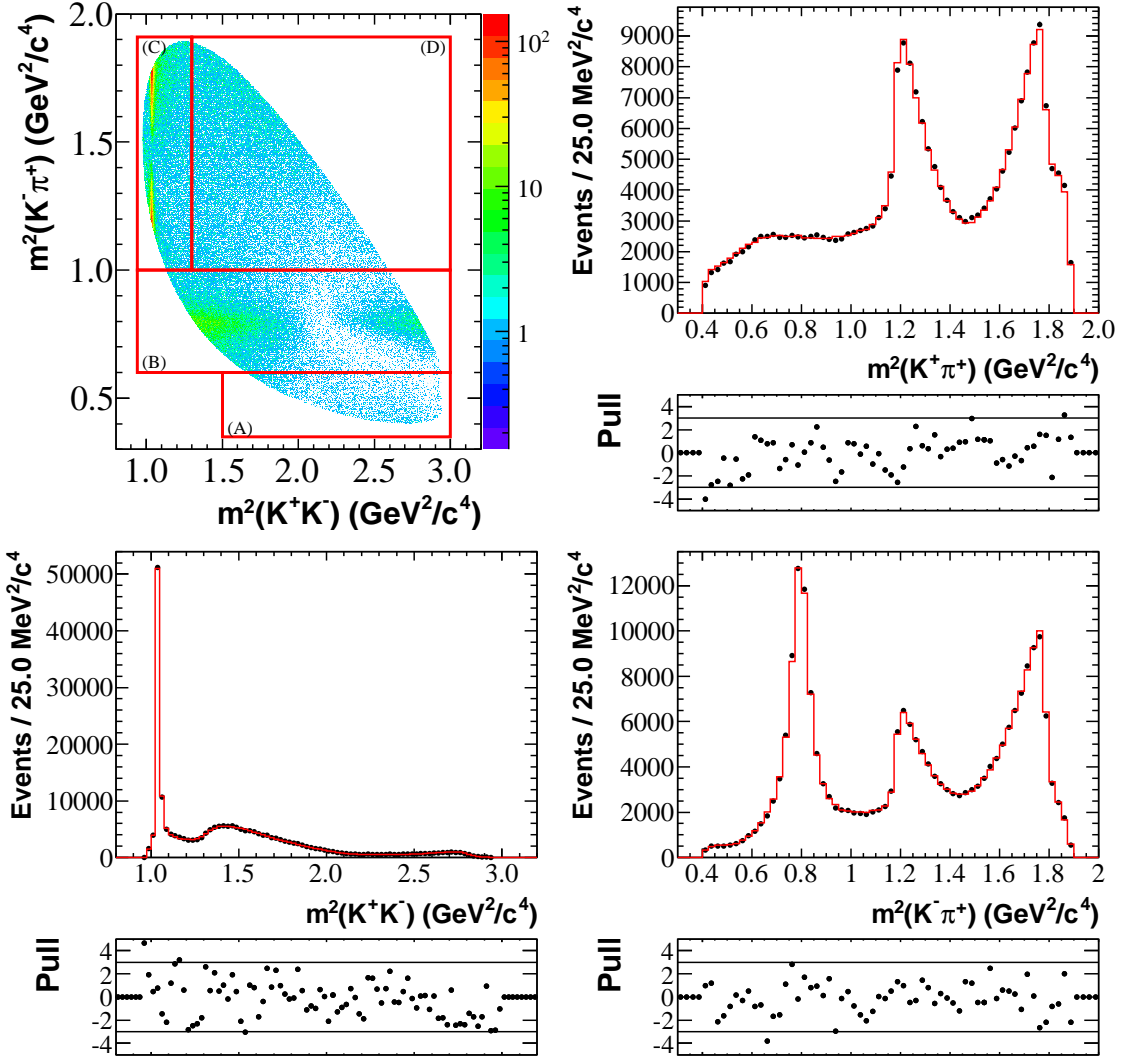}
\vspace{-0.3cm}
\caption{$\Dpm\rightarrow \kk\pi^{\pm}$ Dalitz plot and
  fit projections assuming no {\CPV}, with the regions used for
  model-independent comparisons indicated as boxes. The A/B boundary is
  at $m_{K\pi} = 0.6 \; $GeV$^2/c^4$, the B/C boundary at $m_{K\pi} =
  1.0 \; $GeV$^2/c^4$, and the C/D boundary at $m_{KK} = 1.3 \;
  $GeV$^2/c^4$. In the fit projections, the data are represented by
  points with error bars and the fit results by the histograms. The
  normalized residuals below each projection, defined as
  $(N_{\mathrm{Data}} - N_{\mathrm{MC}})/\sqrt{N_{\mathrm{MC}}}$, lie
  between $\pm5\sigma$.  The horizontal lines correspond to $\pm
  3\sigma$.}
\label{fig:dpfit}
\vspace{-0.7cm}
\end{center}
\end{figure*}

The mass and width values of several resonances, including the \Kres~and
\phires, are determined in the fit (Table~\ref{tab::fit_resparms}). The
$f_0(980)$ resonance is modeled with an effective BW parameterization:
\begin{equation}
A_{f_0(980)} = \frac{1}{m_0^2 - m^2 - im_0\Gamma_0\rho_{KK}},
\end{equation}
determined in the partial-wave analysis of $D_s^+\to K^+K^-\pi^+$
decays~\cite{delAmoSanchez:2011dskkpi}, where $\rho_{KK} = 2p/m$ with
$p$ the momentum of the \Kp in the $K^+K^-$ rest frame, $m_0 = 0.922
\gevcc$, and $\Gamma_0 = 0.24 \gev$. The remaining resonances (defined
as $r\to AB$) are modeled as relativistic BWs:
\begin{equation}
{\mathrm{RBW}}(M_{AB}) = \frac{F_rF_D}{M_r^2 - M_{AB}^2 - i\Gamma_{AB}M_r},
\end{equation}
where $\Gamma_{AB}$ is a function of the mass $M_{AB}$, the momentum
$p_{AB}$ of either daughter in the $AB$ rest frame, the spin of the
resonance, and the resonance width $\Gamma_R$. The form factors $F_r$
and $F_D$ model the underlying quark structure of the parent particle of
the intermediate resonances.  Our model for the \kpi ${\mathcal S}$-wave
term consists of the $\kappa(800)$, the $\bar{K}_0^*(1430)^0$, and a
nonresonant amplitude. Different parameterizations for this
term~\cite{kpipimipwa, delAmoSanchez:2010d0kshh} do not provide a better
description of data. The resulting fit fractions are summarized in
Table~\ref{tab:ff}. A $\chi^2$ value is defined as
\begin{equation}
  \chi^2 = \sum_i^{N_{\mathrm{bins}}} \frac{(N_i - N_{\mathrm{MC}_i})^2}{N_{\mathrm{MC}_i}}
\end{equation}
where $N_{\mathrm{bins}}$ denotes 2209 intervals of variable size. The
$i^{th}$ interval contains $N_i$ events (around 100), and $N_{\mathrm{MC}_i}$
denotes the integral of the Dalitz-plot model within the interval. The goodness-of-fit
$\chi^2/$ndof = 1.21 for ndof $= 2165$.  The distribution of the
data in the Dalitz plot, the projections of the data and the model of
the Dalitz plot variables, and the one-dimensional residuals of the data
and the model, are shown in Fig.~\ref{fig:dpfit}.

\begin{table}[!ht]
\begin{center}
\caption{Fit fractions of the resonant and nonresonant amplitudes in the
  isobar model fit to the data. The uncertainties are statistical.
}
\setlength{\extrarowheight}{2pt}
\begin{tabular}{lc}
\hline 
\hline
Resonance & Fraction ($\%$) \\ 
\hline
$\bar{K}^{*}(892)^{0}$ & 21.15 $\pm$ 0.20 \\
$\phi(1020)$ & 28.42 $\pm$ 0.13 \\
$\bar{K}^{*}_{0}(1430)^{0}$ & 25.32 $\pm$ 2.24 \\
NR & 6.38 $\pm$ 1.82 \\
$\kappa(800)$ & 7.08 $\pm$ 0.63 \\
$a_{0}(1450)^{0}$ & 3.84 $\pm$ 0.69 \\
$f_{0}(980)$ & 2.47 $\pm$ 0.30 \\
$f_{0}(1370)$ & 1.17 $\pm$ 0.21 \\
$\phi(1680)$ & 0.82 $\pm$ 0.12 \\
$\bar{K}^{*}_{1}(1410)$ & 0.47 $\pm$ 0.37 \\
$f_{0}(1500)$ & 0.36 $\pm$ 0.08 \\
$a_{2}(1320)$ & 0.16 $\pm$ 0.03 \\
$f_{2}(1270)$ & 0.13 $\pm$ 0.03 \\
$\bar{K}^{*}_{2}(1430)$ & 0.06 $\pm$ 0.02\\
$\bar{K}^{*}(1680)$ & 0.05 $\pm$ 0.16 \\
$f_{0}(1710)$ & 0.04 $\pm$ 0.03\\
$f'_{2}(1525)$ & 0.02 $\pm$ 0.01\\
\hline
Sum & 97.92 $\pm$ 3.09 \\
\hline
\hline  
\end{tabular} 
\label{tab:ff}
\end{center}
\end{table}

\begin{figure*}[!tb]
\begin{center}
\includegraphics[width=0.3\textwidth]{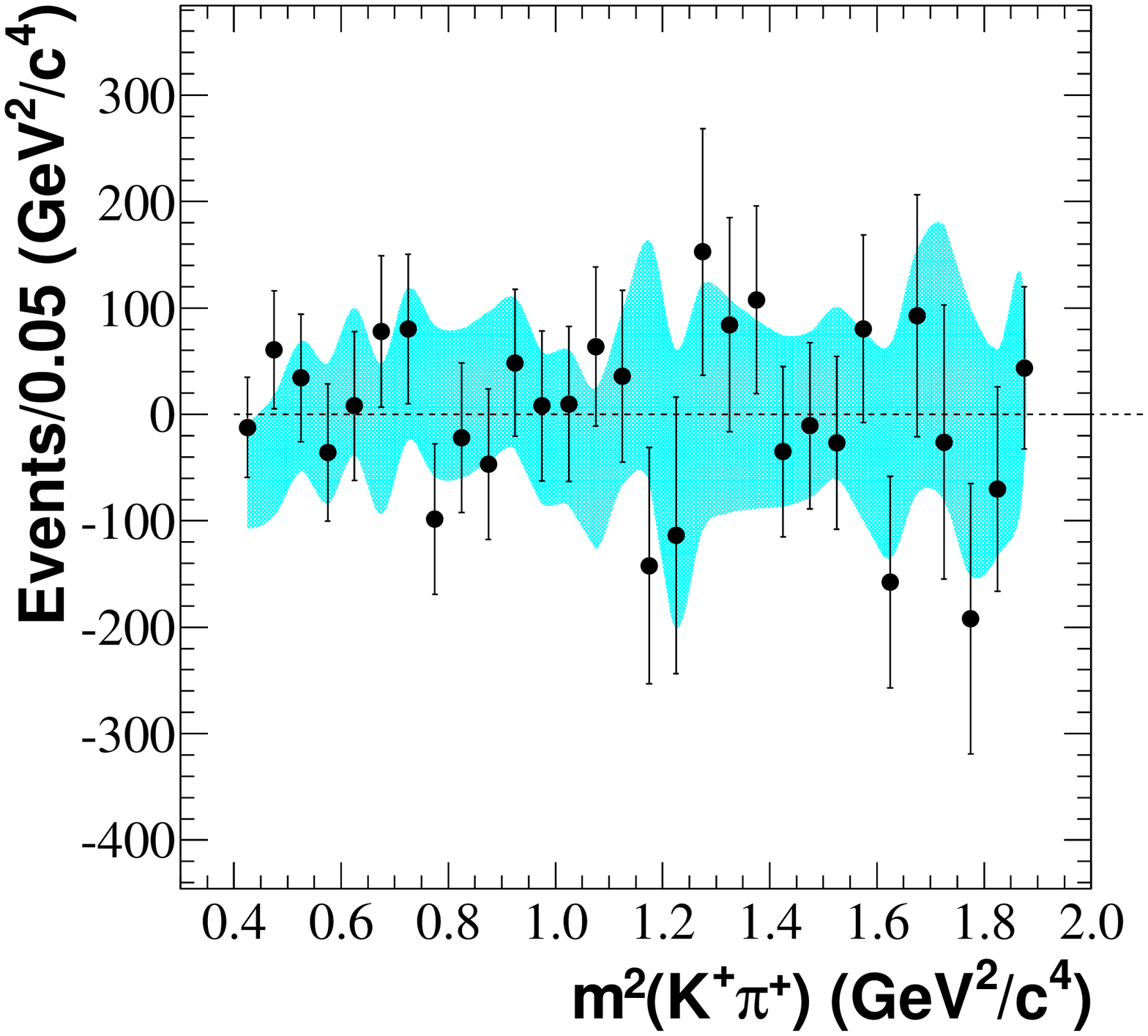}
\includegraphics[width=0.3\textwidth]{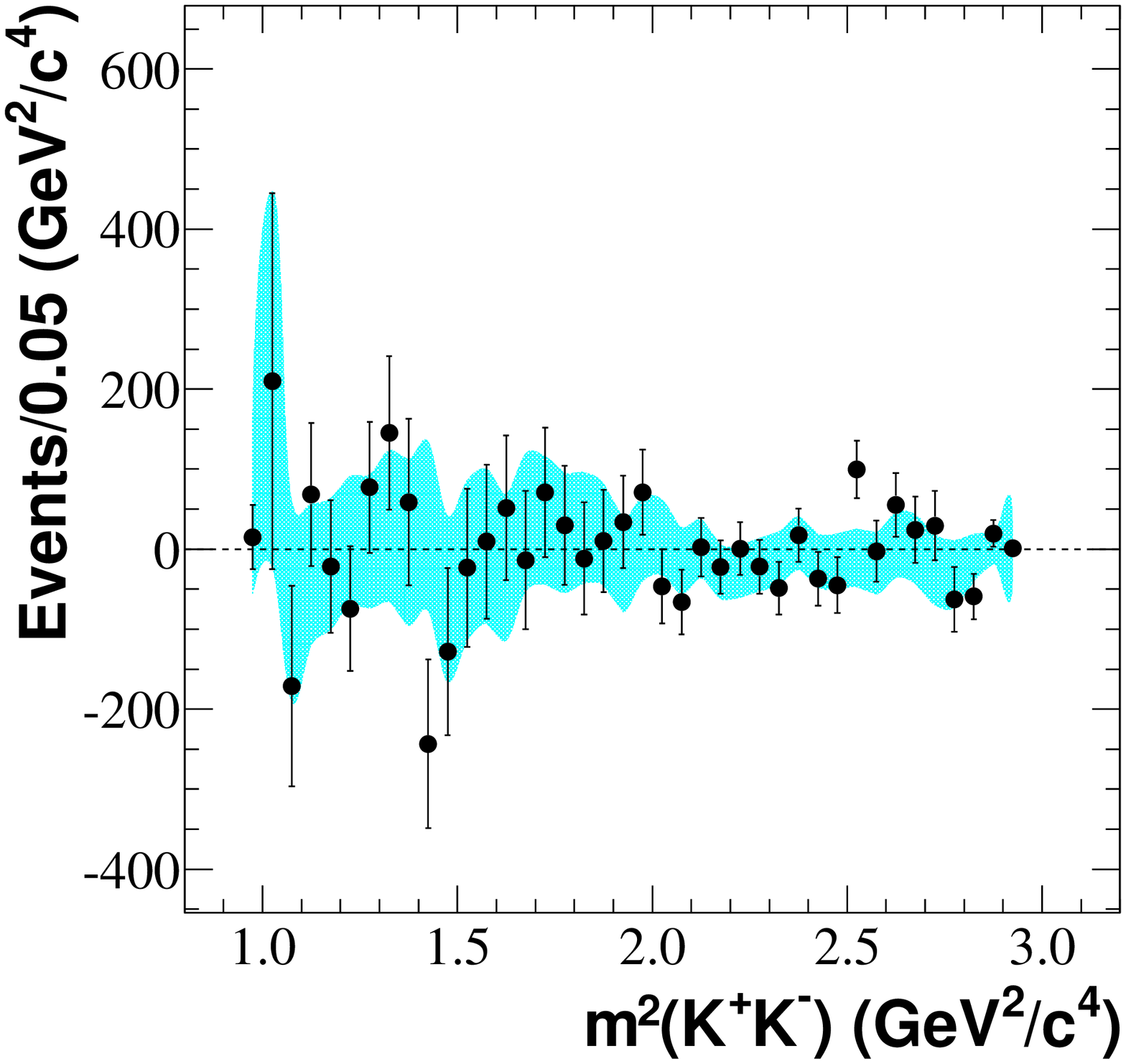}
\includegraphics[width=0.3\textwidth]{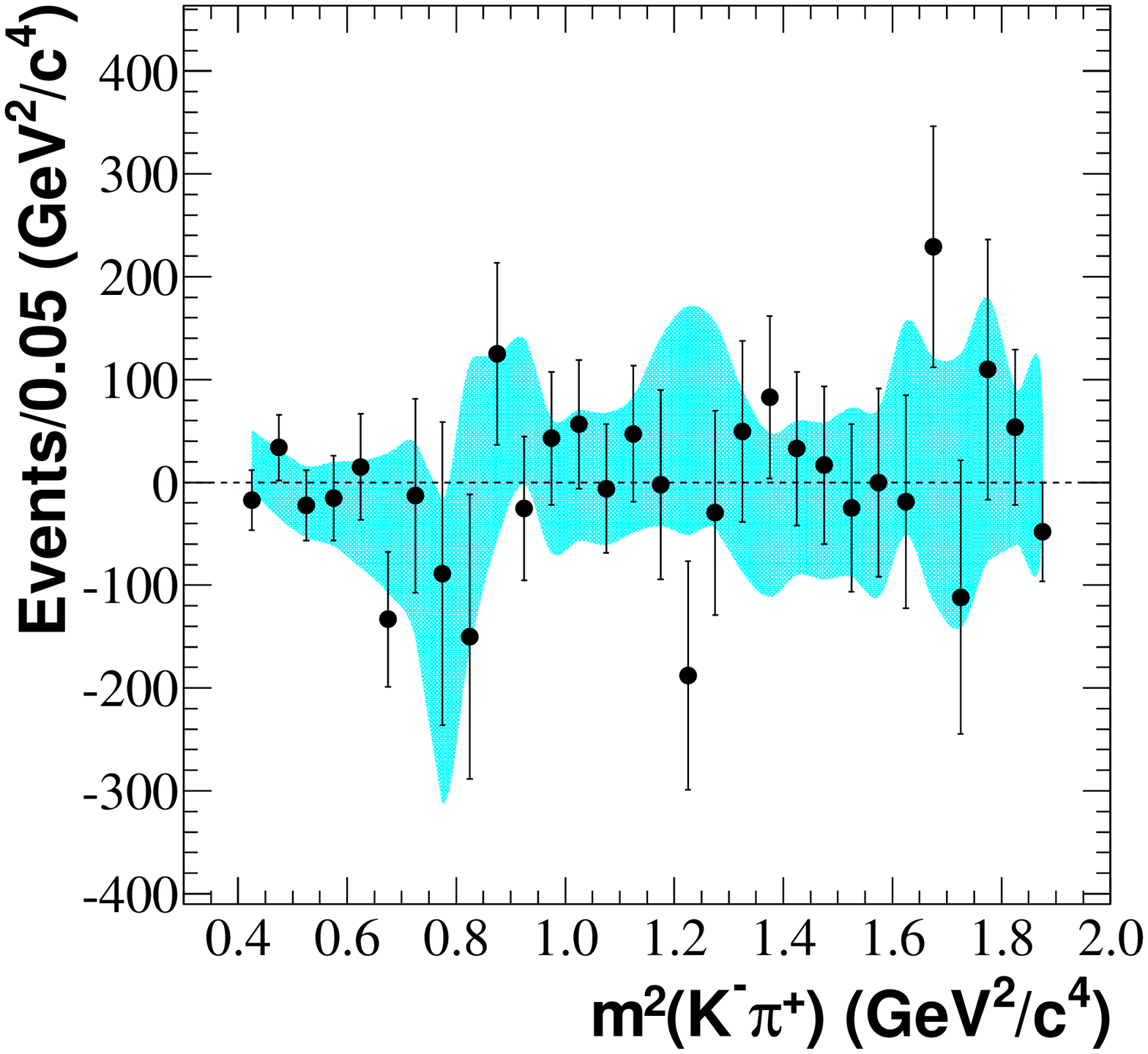}
\vspace{-0.3cm}
\caption{The difference beween the \Dp and \Dm Dalitz
  plot projections of data (points) and of the fit (cyan band). The
  width of the band represents the $\pm1$ standard deviation statistical
  uncertainty expected for the size of our data sample.}
\label{fig:cpv}
\vspace{-0.7cm}
\end{center}
\end{figure*}

To allow for the possibility of {\CPV} in the decay, resonances with a
fit fraction of at least 1$\%$ (see Table ~\ref{tab:ff}) are permitted
to have different \Dp and \Dm magnitudes and phase angles in the decay
amplitudes (${\mathcal A}$ or $\bar{{\mathcal A}}$). A simultaneous fit is performed
to the \Dp and \Dm data, where each
resonance has four parameters: ${\mathcal M_r},\phi_r, r_{CP}$, and
$\Delta\phi_{CP}$. The \CP-violating parameters are $r_{CP} =
\frac{|{\mathcal M}_r|^2 - |\bar{{\mathcal M_r}}|^2}{|{\mathcal M}_r|^2
+ |\bar{{\mathcal M_r}}|^2}$ and $\Delta\phi_{CP} = \phi_r -
\bar{\phi}_r$. In the case of ${\mathcal S}$-wave resonances in the
$K^+K^-$ system, which make only small contributions to the model,
used instead are the Cartesian-form of the \CP\ parameters, $\Delta x$ and
$\Delta y$, to parameterize the amplitudes and asymmetries. This choice
of parameterization removes or eliminates technical problems with the
fit. For these resonances, the parameters
$x_r(D^\pm) = x_r \pm \Delta x_r/2$ and $y_r(D^\pm) = y_r \pm \Delta
y_r/2$ are introduced. The masses and widths determined in the initial fit (shown in
Table~\ref{tab::fit_resparms}) are fixed, while the remaining parameters
are determined in the fit. Table~\ref{tab:dpmodel} summarizes the \CP
asymmetries, \ie, either the polar-form pair $(r_{CP}, \Delta\phi_{CP})$
or the Cartesian pair $(\Delta x_r, \Delta y_r)$.  Figure~\ref{fig:cpv}
shows the difference between the Dalitz-plot projections of the \Dp and
\Dm decays, for both the data and the fit. It is
evident from the figure that both the charge asymmetry of the data and
fit are consistent with zero and with each other.

\begin{table*}
\centering
\setlength{\extrarowheight}{5pt}
\caption{\CP-violating parameters from the simultaneous Dalitz plot fit. The first
  uncertainties are statistical and the second are systematic.}
\begin{tabular}{lccc}
\hline 
\hline
Resonance & $r_{CP}$ ($\%$) & $\Delta\phi\;(^{\circ})$\\ 
\hline
$\bar{K}^{*}(892)^{0}$          & 0. (FIXED)                                    & 0. (FIXED)                            \\
$\phi(1020)$                    & $0.35^{+0.82}_{-0.82} \pm 0.60$    & $7.43^{+3.55}_{-3.50} \pm 2.35$    \\ 
$\bar{K}^{*}_{0}(1430)^{0}$     & $-9.40^{+5.65}_{-5.36} \pm 4.42$   & $-6.11^{+3.29}_{-3.24} \pm 1.39$   \\
NR                              & $-14.30^{+11.67}_{-12.57} \pm 5.98$ & $-2.56^{+7.01}_{-6.17} \pm 8.91$  \\
$\kappa(800)$                   & $2.00^{+5.09}_{-4.96} \pm 1.85$    & $2.10^{+2.42}_{-2.45} \pm 1.01$    \\ 
$a_{0}(1450)^{0}$               & $5.07^{+6.86}_{-6.54} \pm 9.39$            & $4.00^{+4.04}_{-3.96} \pm 3.83$ \\
\hline
		& $\Delta x$                            & $\Delta y$                    \\              
$f_{0}(980)$                    & $-0.199^{+0.106}_{-0.110} \pm 0.084$ & $-0.231^{+0.100}_{-0.105} \pm 0.079$ \\
$f_{0}(1370)$                   & $0.019^{+0.049}_{-0.048} \pm 0.022$        & $-0.0045^{+0.037}_{-0.039} \pm 0.016$ \\

		    \hline
\hline
\end{tabular} 
\label{tab:dpmodel}
\end{table*}

In summary, the \babar~and Belle collaborations have studied the SCS \Dkkpi decay using complimentary
analysis techniques to measure \CP violating asymmetries and search for new physics beyond the SM. The Belle
measurement probed for \CPV in the dominant quasi two-body SCS decay mode $\Dp\to\phi\pi^+$ and CF mode
$D_s^+\to\phi\pi^+$, resulting in a precise measurement with small systematic uncertainties. The \babar~measurement 
took advantage of the Dalitz plot decay to probe for \CPV in all regions of the phase-space, making use of both model-dependent
and model-independent techniques. No \CPV is observed in either measurement. Further studies to improve the description of the 
Dalitz plot may provide a deeper understanding of the dynamics in three-body decays and \CPV in charm decays. 

\Acknowledgements
I am grateful to Milind V. Purohit for his advice and collaboration over the years, and for the 
extraordinary contributions of our \pep2\ colleagues in
achieving the excellent luminosity and machine conditions
that have made much of this work possible. This work is supported by the
US Department of Energy.

\end{document}